# Percolation Model Explaining Both Unipolar Memory and Threshold Resistance Switchings in NiO Film

S. H. Chang,[1] J. S. Lee,[2] S. C. Chae,[1] S. B. Lee,[1] C. Liu,[1] B. Kahng,[2] D.-W. Kim,[3] and T. W. Noh[1,*]

[1]*ReCOE & FPRD, Department of Physics and Astronomy, Seoul National University, Seoul 151-747, Korea*

[2]*Department of Physics and Astronomy, Seoul National University, Seoul 151-747, Korea*

[3]*Division of Nano Sciences and Department of Physics, Ewha Womans University, Seoul 120-750, Korea*

We observed two types of unipolar resistance switching (RS) in NiO film: memory RS at low temperature and threshold RS at high temperature. We explain these phenomena using a bond percolation model that describes the forming and rupturing of conducting filaments. Assuming Joule heating and thermal dissipation processes in the bonds, we explain how both RS types could occur and be controlled by temperature. We show that these unipolar RS are closely related and can be explained by a simple unified percolation picture.

———————————————

*E-mail: twnoh@snu.ac.kr

In the early 1960s, some binary oxides, such as NiO [1] and $Nb_2O_5$ [2], were reported to have bistable resistance states that could be switched by applying an electrical bias. These reversible resistance switching (RS) phenomena have also been observed in many other materials [3–18]. Recently, interest in these intriguing phenomena has been renewed due to their possible application to nonvolatile memory as resistance random access memory (RRAM) [11]. However, because the details of RS behavior have varied from sample to sample, there is confusion on how to even classify these extensively observed RS phenomena.

There are at least three reported RS types: (i) unipolar memory RS [3–12], (ii) unipolar threshold RS [11–13], and (iii) bipolar memory RS [14–18]; their characteristic current-voltage (*I-V*) curves are shown in Figs. 1(a)–(c). In unipolar RS, the *I-V* curves are symmetric upon inversion of the electric bias polarity; see Figs. 1(a) and (b). Figure 1(a) shows RS from a low resistance state (LRS) to a high resistance state (HRS) at a certain voltage and another RS from a HRS to a LRS at a higher voltage. The latter process is usually controlled by setting a limit to the total amount of current flow, the compliance current ($I_{comp}$), to prevent a complete dielectric breakdown. As these two states are stable at $V = 0$, they can be used in a nonvolatile memory device. In Fig. 1(b), however, there is only one stable resistance state; without applying external bias, the lower resistance state does not remain stable. Conversely, as shown in Fig. 1(c), the *I-V* curve of bipolar RS is asymmetric with polarity switching; at a positive bias, a LRS will change into a HRS, and vice versa.

Considerable theoretical and experimental efforts have been expended to understand why RS occurs. For unipolar memory RS, numerous models have been proposed, including those based on charge trapping and detrapping of intragap states [3], oxygen electromigration [4,5], and Mott metal-insulator transitions at interfaces [6]. Suggested models for unipolar threshold RS include chemical stoichiometry [11], formation of conducting filaments [12], and double injection of charge carriers [13]. There have been even more models proposed for bipolar memory RS [14–18]. In spite of such extensive efforts, there has been little consensus on the basic mechanism of any of the RS phenomena, and whether any RS is related to any other.

In this Letter, we report that a NiO film can show unipolar memory RS at low temperature and unipolar threshold RS at high temperature. (As we will deal only with unipolar RS phenomena, we will omit the word 'unipolar' from now on.) We attribute both RS to forming and rupturing of conducting filaments, where the rupturing process occurs through a subtle balance between Joule heating and thermal dissipation in the bonds. Based on this unified picture, we developed a bond percolation model to explain (1) how both memory and threshold RS can occur in the same sample, and (2) why the RS type can be controlled by thermal cycling.

We prepared a Pt/Ti/SiO$_2$/Si substrate having a Pt bottom electrode layer of 30 nm. We then deposited a Ni film using e-beam evaporation, and oxidized it into a NiO layer at 450°C in ambient air for 1 hr. The resultant NiO film was approximately 60 nm thick. We followed this by evaporating a Au(30 nm)/Pt(10 nm) top electrode layer to make a series of Pt/NiO/Pt capacitors. The resistance switching characteristics of these Pt/NiO/Pt capacitors were examined by measuring *I-V* curves at temperatures ($T$) between 80 and 300 K. Further details of the sample preparation methods and *I-V* measurements are presented elsewhere [10].

At low *T*, our NiO film showed a typical memory RS. Figure 2(a) shows the NiO film *I-V* curves at 118 K. Under an applied external bias *V* of about 6.2 V, the pristine sample changed its resistance state, marked on the graph as "Forming." It then entered into a LRS. When we started to increase *V* from 0 V again, it suddenly changed into a HRS at $V \approx 2.0$ V, marked as "Reset." This HRS was stable even without external bias, and changed into a LRS again at $V \approx 4.5$ V, marked as "Set." For both the forming and set processes, $I_{comp}$ was about 10 mA.

We could control the RS type of our NiO film by changing $T$. When we increased $T$ to 300 K, the $I$-$V$ curve was typical of threshold RS, as shown in Fig. 2(b); cooling the sample to 80 K, caused the reappearance of memory RS, Fig. 2(c); and increasing $T$ to 300 K, resulted in threshold RS again, Fig. 2(d). These thermal cycles were repeated several times, and we still observed the $T$-dependent RS changes. These results show that both memory and threshold RS can be obtained from one sample by repeated thermal cycling, indicating that these two RS are closely connected to each other.

The detailed behaviors of threshold RS, shown in the inset of Fig. 2(d), show how we can relate these two types of RS. The state with the lower resistance, generated by applying $V$, is not stable without an external bias, so it returns to the original higher resistance state with a decrease of $V$. This instability in the lower resistance state could be the main difference between observed memory and threshold RS.

To obtain further insight, we plotted the $T$-dependent values of the conductance, $G(V = 0.2$ V$)$, of the NiO film at $V \approx 0.2$ V. Figure 3 shows the average values of $G(V = 0.2$ V$)$, measured on several NiO capacitors, for various values of $T$. In memory RS, the LRS $G(V = 0.2$ V$)$ is weakly $T$-dependent and shows a metallic behavior. On the other hand, the HRS $G(V = 0.2$ V$)$ shows an insulating behavior and follows $G \sim \exp(-\phi/k_B T)$, with an activation energy of $\phi = 75.8$ meV. This $T$-dependence is consistent with earlier observations for memory RS [7]. Note that the $G$ values for the stable state in threshold RS fall approximately on the same $G \sim \exp(-\phi/k_B T)$ curve as for the HRS in memory RS. If we assume that the corresponding LRS becomes unstable in threshold RS, namely at a higher $T$, we might be able to explain both RS of our NiO film using a simple unified picture.

Quite recently, we suggested a new type of bond percolation model, called the "random circuit breaker (RCB) network model," for unipolar memory RS [9]. There are earlier reports on the existence of conducting filamentary paths in samples with memory RS [1,4,5]. We postulated that the forming and rupturing of such conducting filaments could be represented by the turning-on and -off of circuit breakers. We then treated the transport in our thin films as a bond percolation problem with an M×N lattice network composed of circuit breakers, as shown in Fig. 4(a). Each circuit breaker is in either an on- or off-state, with corresponding resistance values $r_l$ and $r_h$ [$r_h \gg r_l$]. The thick (red) and thin (blue) lines in Fig. 4(a) represent these bond on- and off-states.

In the original RCB network model, the switchability of the bond resistance value was assumed to depend only on the bias voltage, $\Delta v$, applied to each circuit breaker. As shown in Fig. 4(b), the on-state changes into the off-state when $\Delta v > v_{\text{off}}$, and the off-state returns to the on-state when $\Delta v > v_{\text{on}}$ [$v_{\text{on}} > v_{\text{off}}$]. Note that $\Delta v$ depends on the distribution of the on-/off-states of the circuit breakers, whose initial values are randomly chosen as a given fraction in the simulations. By increasing $V$, more bonds are turned on. When a percolating cluster of on-state circuit breakers is formed through the RCB network, the sample will be in a LRS. Then, when such a cluster becomes broken by turning off one or several circuit breakers, it will be in a HRS. This RCB network model was successfully applied to explain most of the experimental observations of memory RS, including the basic operations (shown in Fig. 1(a)), *I-V* duality, universal resistance jump, and the switching voltage distribution [9].

To explain the *T*-dependent RS change in our NiO film, we extended the RCB network model by assuming that the turning-off process could be affected by a bond's local temperature $T_{\text{loc}}$ change due to Joule heating and thermal dissipation. Note that the current flow in the RCB network is highly inhomogeneous. Since $r_h \gg r_l$, most of the current will flow through the largest cluster of on-state bonds. According to the percolation theory, just above the percolation threshold, most of the current will flow in a single bond, called the "hottest bond" [19], and its Joule-heating effects could become very important during the reset process. To evaluate $T_{loc}$, we made the simple assumption that each bond can exchange heat with a thermal bath of temperature $T_b$, as shown in Fig. 4(c) [20]. We also extended the turning-off condition of a bond to include thermal effects. Then Joule heating due to the flow of current $i$ and the heat exchange process will compete, so

$$c\, dT_{loc}/dt = ri^2 - a(T_{loc} - T_b), \qquad (1)$$

where $c$, $r$, and $a$ are the heat capacitance, resistance, and thermal conductance of each bond, respectively. As shown in Fig. 4(c), the circuit breaker changes its resistance from $r_l$ to $r_h$ when $T_{loc}$ becomes higher than a certain threshold temperature $T_c$.

Using this modified RCB model, we obtained *I-V* curves for two thermal bath temperatures, i.e., $T_b/T_c = 0.3$ and 0.75. In our simulations, we used a 50 × 20 two-dimensional square lattice. We performed simulations with a pristine state, where 0.5% of the circuit breakers were randomly chosen to be in an on-state and $T_{loc}$ for all bonds equaled $T_b$. We increased *V* in steps from zero, assuming that the external bias at each step was applied for fixed time duration, $t_d$, corresponding to an experimental measurement time. At $t = 0$, the $\Delta v$ distribution was evaluated by solving Laplace's equation. If $\Delta v$ of at least one off-state bond satisfied the switching condition of $\Delta v > v_{on}$, we changed $r_h$ to $r_l$, reevaluated the $\Delta v$ distribution, and rechecked the switching condition. This process was iterated until a stable $\Delta v$ distribution was reached. After that, we calculated the time-dependent changes of $T_{loc}$ of each bond using Eq. (1) until $t$ reached $t_d$. If $T_{loc}$ of an on-state circuit breaker reached $T_c$ when $t < t_d$, we changed $r_l$ to $r_h$ and recalculated a stable $\Delta v$ distribution. If $t$ reached $t_d$ without switching, we increased *V* by $\Delta V$ ($\Delta V = 0.05$ V) and repeated the whole procedure until the current flowing through the network reached $I_{comp}$. When the forming operation occurred, we set *V* to zero immediately, and then calculated the time-dependent changes of $T_{loc}$ and kept updating the on-/off-states using the $T_{loc} >$

$T_c$ switching condition. Once a stable state was reached, we started to increase $V$ from zero again to obtain the next *I-V* curve.

Figures 5(a) and (b) show simulated *I-V* curves for $T_b/T_c$ = 0.3 and 0.75, respectively. We set $v_{on}$ = 1.0 and $I_{comp}$ = 0.5 in arbitrary units. For the turned-off and -on states of the circuit breaker, we adopted values of $c$ and $a$ from Ref. 1. For lower $T_b$ simulations, we used $r_h/r_l$ = 10000 [21]. As shown in Fig. 5(a), the results show typical memory RS. On the other hand, for higher $T_b$, we used $r_h/r_l$ = 100, representing the observed change in resistance from 300 K to 118 K shown in Fig. 3. The obtained *I-V* curves in Fig. 5(b) represent those of threshold RS. For completeness, we also obtained an *I-V* curve while we gradually decreased $V$ after it reached the formed state. As shown in the inset of Fig. 5(b), these results agree quite well with our observed *I-V* curves for the threshold RS shown in Fig. 2(d). This simulation demonstrates that both memory and threshold RS can be described by the unified percolation model.

The RS type could be determined by a competition between Joule heating and thermal dissipation processes, especially in the hottest bond. When $T_{loc}$ approaches $T_c$, Eq. (1) becomes

$$\frac{dT_{loc}}{T_c - T_b} \sim [\frac{1}{\tau_1} - \frac{1}{\tau_2}]dt, \qquad (2)$$

where $\tau_1 = c(T_c - T_b)/(v_{on}^2/r_h)$ and $\tau_2 = c/a$. Note that the first and second terms come from Joule heating and thermal dissipation processes, respectively [22]. If $[1/\tau_1 - 1/\tau_2]$ becomes positive, i.e., $\tau_1 < \tau_2$, $T_{loc}$ increases above $T_c$ as time progresses. The conducting filament will then become unstable. On the other hand, if $[1/\tau_1 - 1/\tau_2]$ becomes negative, i.e., $\tau_1 > \tau_2$, $T_{loc}$ cannot reach $T_c$ as time passes, leaving the filament stable, as in memory RS. In our experimental studies, temperature can induce a change of $\tau_1$ (due to $T_b$ and $r_h$ changes), resulting in a change in the RS type [23].

In summary, our study indicates that memory and threshold resistance switching for NiO films could come from the same origins, i.e., forming and rupturing of conducting filamentary paths. The stability of such paths will be determined by the competition between Joule heating and thermal dissipation processes. Using a simple unified picture, based on a percolation model, we explained how both resistance switching types can occur in the same sample and can be controlled by temperature.

This work was supported financially by the Creative Research Initiatives (Functionally Integrated Oxide Heterostructure) of the Ministry of Science and Technology (MOST) and the Korean Science and Engineering Foundation (KOSEF).

[1] J. F. Gibbons and W. E. Beadle, Solid-State Electron. **7**, 785 (1964).

[2] W. Hiatt and T. W. Hickmott, Appl. Phys. Lett. **6**, 106 (1965).

[3] J. G. Simmons and R. R. Verderber, Proc. Royal Society A. **301,** 77 (1967).

[4] Y. Sato, K. Kinoshita, M. Aoki, and Y. Sugiyama, Appl. Phys. Lett. **90**, 033503 (2007).

[5] K. M. Kim, B. J. Choi, Y. C. Shin, S. Choi, and C. S. Hwang, Appl. Phys. Lett. **91**, 012907 (2007).

[6] M. J. Sánchez, M. J. Rozenberg, and I. H. Inoue, Appl. Phys. Lett. **91**, 252101 (2007).

[7] K. Jung, H. Seo, Y. Kim, H. Im, J. P. Hong, J.-W. Park, and J.-K. Lee, Appl. Phys. Lett. **90**, 052104 (2007).

[8] I. H. Inoue, S. Yasuda, H. Akinaga, and H. Takagi, Phys. Rev. B **77**, 035105 (2008).

[9] S. C. Chae, J. S. Lee, S. Kim, S. B. Lee, S. H. Chang, C. Liu, B. Kahng, H. Shin, D.-W. Kim, C. U. Jung, S. Seo, M.-J. Lee, and T. W. Noh, Adv. Mater. **20**, 1154 (2008).

[10] S. H. Chang, S. C. Chae, S. B. Lee, C. Liu, J. S. Lee, B. Kahng, J. H. Jang, M. Y. Kim, D.-W. Kim, C. U. Jung, and T. W. Noh, arXiv: 0802.3739v1.

[11] S. Seo, M. J. Lee, D. H. Seo, E. J. Jeoung, D.-S. Suh, Y. S. Joung, I. K. Yoo, I. R. Hwang, S. H. Kim, I. S. Byun, J.-S. Kim, J. S. Choi, and B. H. Park, Appl. Phys. Lett. **85**, 5655 (2004).

[12] G. Dearnaley, A. Stoneham, and D. V. Morgan, Rep. Prog. Phys. **33**, 1129 (1970).

[13] H. K. Henisch, E. A. Fagen, and S. R. Ovshinsky, J. Non-Cryst. Solids **4**, 538 (1970).

[14] S. Tsui, A. Baikalov, J. Cmaidalka, Y. Y. Sun, Y. Q. Wang, Y. Y. Xue, C. W. Chu, L. Chen, and A. J. Jacobson, Appl. Phys. Lett. **85**, 317 (2004).

[15] M. J. Rozenberg, I. H. Inoue, and M. J. Sánchez, Phys. Rev. Lett. **92**, 178302 (2004).

[16] A. Sawa, T. Fujii, M. Kawasaki, and Y. Tokura, Appl. Phys. Lett. **85**, 4073 (2004).

[17] Y. B. Nian, J. Strozier, N. J. Wu, X. Chen, and A. Ignatiev, Phys. Rev. Lett. **98**, 146403 (2007).

[18] R. Waser and M. Aono, Nat. Mater. **6**, 833 (2007), and references therein.

[19] G. G. Batrouni, B. Kahng, and S. Redner, J. Phys. A: Math. Gen. **21**, L23 (1988).

[20] D. Sornette and C. Vanneste, Phys. Rev. Lett. **68**, 612 (1992).

[21] In reality, an increase in the local temperature of a conducting filament will also increase its resistance value. However, for simplicity, we neglected this effect in this simulation.

[22] These two time constants related to the heat generation and dissipation will become important in determining the actual reset programming speed for RRAM.

[23] We also observed that the RS type could be changed by altering the thermal conduction of the thermal heat conduction channels, i.e., a change of $\tau_2$ value. Refer to Ref. 10.

**Figure Captions**

FIG. 1. Schematic diagrams of *I-V* curves for three resistance switching (RS) phenomena: (a) unipolar memory RS, (b) unipolar threshold RS, and (c) bipolar memory RS. The dashed lines on the unipolar RS represent the setting of the compliance current, $I_{comp}$.

FIG. 2 (color online). Temperature (*T*)-dependent changes of RS type in a NiO film. (a), (b), (c), and (d) show the *I-V* curves successively measured at 118 K, 300 K, 80 K and 300 K. Note that memory and threshold RS behaviors are reversible during repeated thermal cycling. The inset in Fig. 2(d) shows details on a linear *y*-scale for the resistance changes, marked with the dashed area.

FIG. 3 (color online). *T*-dependent conductance *G* values at a bias voltage of 0.2 V for the NiO film. There are bistable states for $T \leq 156$ K, i.e., the region with memory RS. However, there is only one stable state for $T \geq 225$ K, i.e., the region with threshold RS.

FIG. 4 (color online). (a) Schematic diagram of a RCB network composed of circuit breakers. (b) The switching rules for turning-on and -off operations in the original RCB network model [Ref. 9]. Note that both switching rules are governed by electric fields. (c) The newly assigned switching rules in this paper. Note that each circuit breaker is assumed to be in thermal contact with a thermal bath of $T_b$. When its local temperature $T_{loc}$ becomes higher than $T_c$, it will change its resistance from $r_l$ to $r_h$ (a thermally driven process). On the other hand, if the applied bias becomes larger than $v_{on}$, the state with $r_h$ will change into $r_l$, independent of $T_{loc}$.

FIG. 5 (color online). Simulation results for *I-V* curves for the cases of $T_b/T_c$ values of (a) 0.3 and (b) 0.75. The former and latter correspond to memory and threshold RS, respectively. The inset in Fig. 5(b) shows the simulation result when *V* is decreased gradually after reaching the formed state.

Figure 1

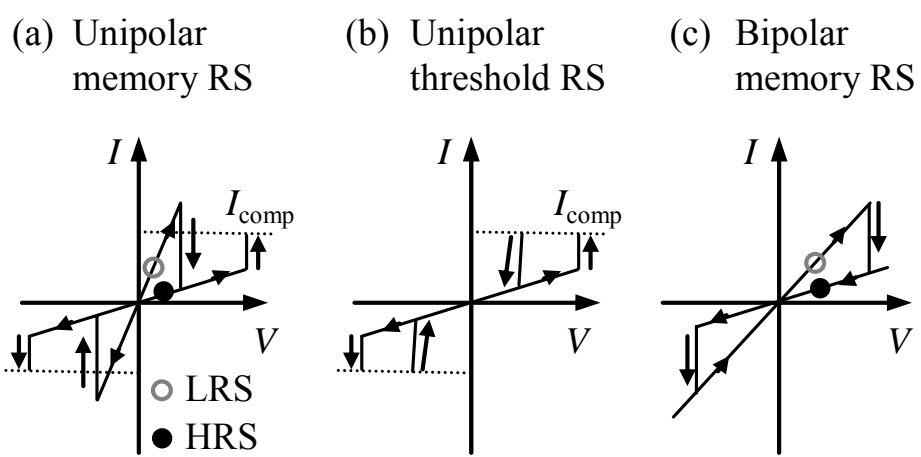

(a) Unipolar memory RS
(b) Unipolar threshold RS
(c) Bipolar memory RS

○ LRS
● HRS

S. H. Chang *et al.*

Figure 2

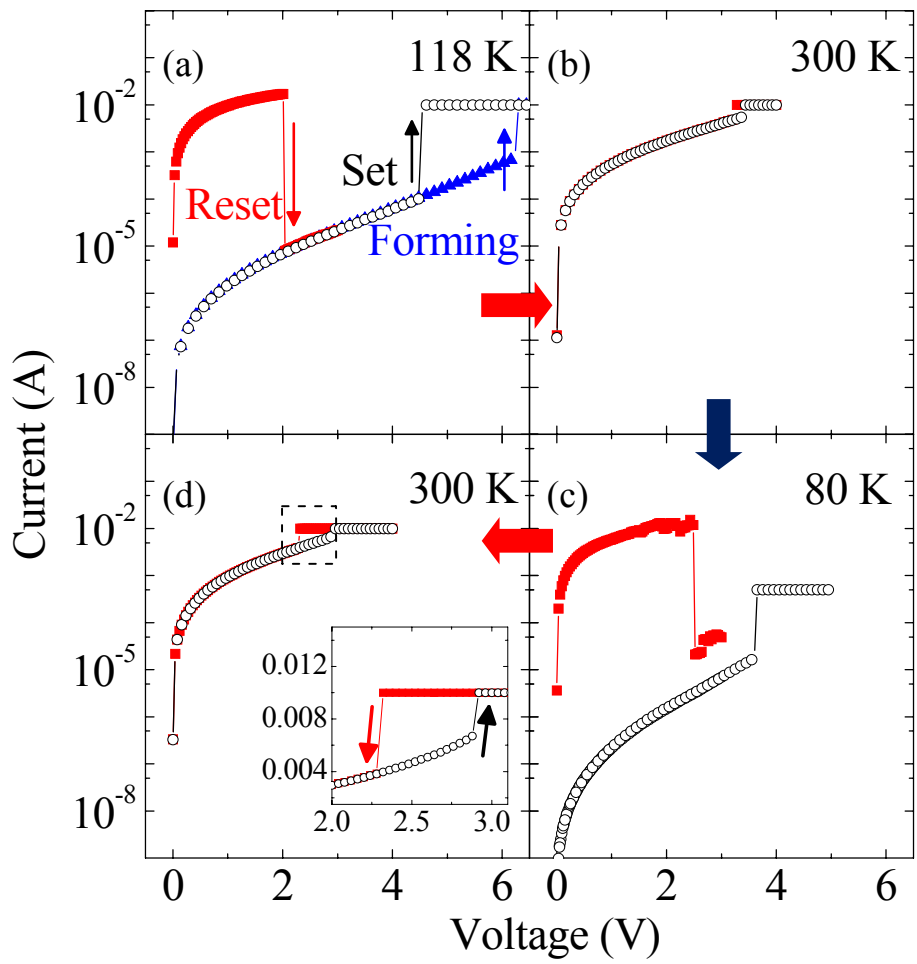

S. H. Chang *et al.*

Figure 3

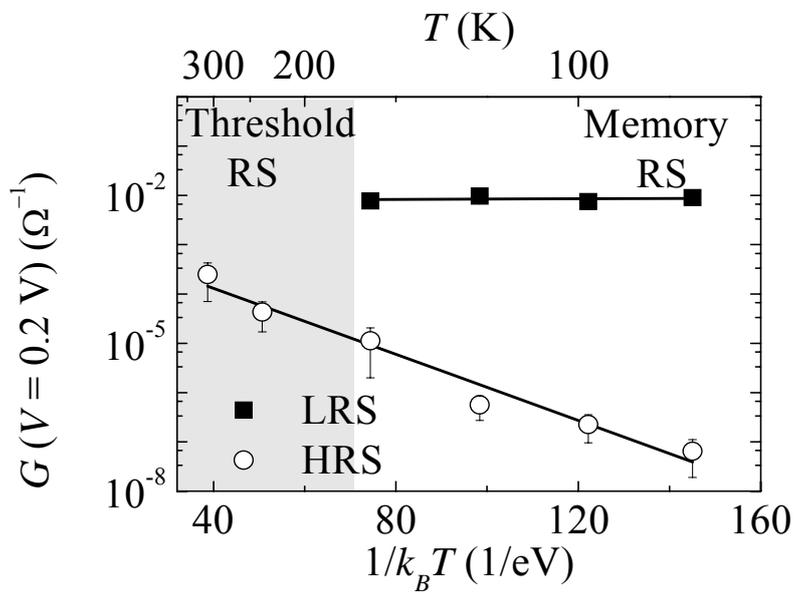

S. H. Chang *et al.*

Figure 4

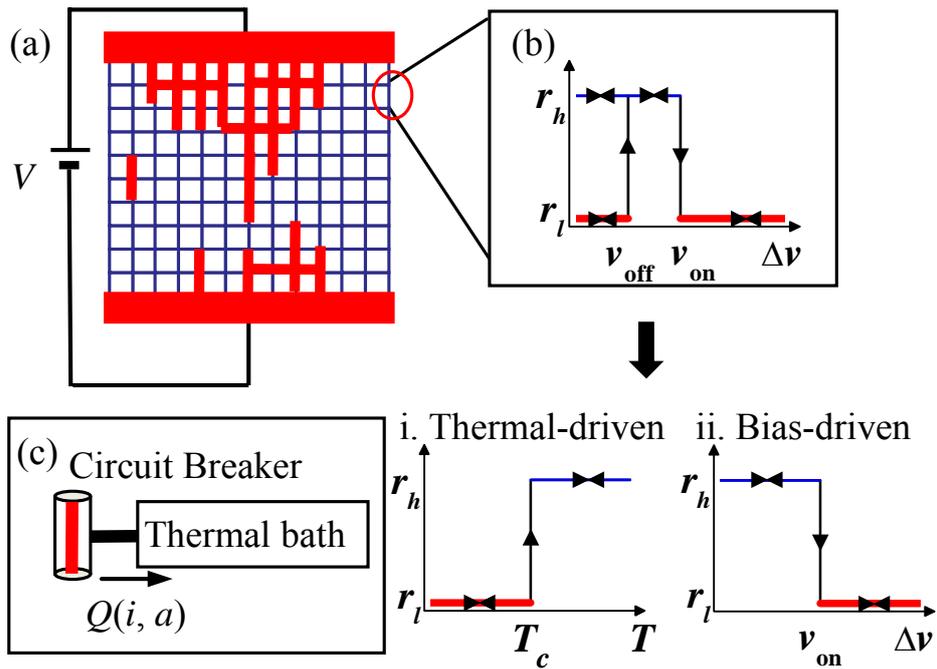

S. H. Chang *et al.*

Figure 5

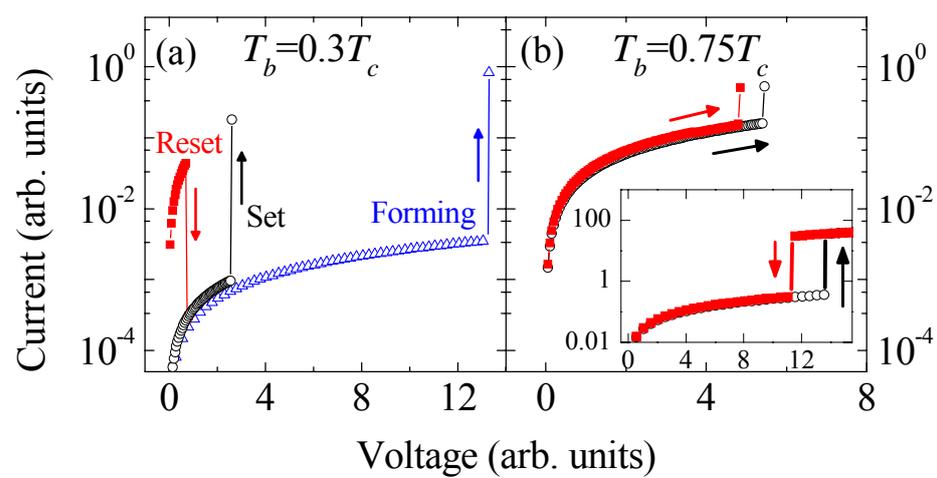

S. H. Chang *et al*.